\documentclass[11pt,a4paper]{amsart}
\usepackage{t1enc}
\usepackage[latin1]{inputenc}
\usepackage[english]{babel}
\normalfont
\usepackage{amsmath}
\usepackage{yfonts}
\usepackage{bbm}
\usepackage{bm}
\usepackage{wasysym}
\usepackage{amssymb}
\usepackage{mathrsfs}
\usepackage{hyperref}

\newcommand{\be}[0]{\begin{equation}}
\newcommand{\ee}[0]{\end{equation}}
\begin{document}

\title{Global anomalies and chiral $p$-forms}

\author{Samuel Monnier}

\address{Institut für Mathematik\\
Mathematisch-naturwissenschaftliche Fakultät\\
Universität Zürich\\
Winterthurerstrasse 190\\
8057 Zürich\\
Switzerland}

\email{samuel.monnier@gmail.com}

\thanks{Research supported in part by SNF Grant No. 200020-131813/1}

\begin{abstract}
This is a short review of the papers \href{http://arxiv.org/abs/1109.2904}{arXiv:1109.2904} and  \href{http://arxiv.org/abs/1110.4639}{arXiv:1110.4639}. After a reminder about local and global gravitational anomalies, we sketch the derivation of the global gravitational anomaly of the self-dual field theory (chiral $p$-forms). We then show that ``cohomological'' type IIB supergravity is free of global gravitational anomalies on all ten-dimensional spin manifolds.
\end{abstract}

\maketitle


\section{Introduction and motivation}

One of the most outstanding problem of modern theoretical physics is to find a fully general and consistent theory of quantum gravity. Unfortunately, in strong contrast to the situation which was prevalent during the development of quantum field theory\footnote{See for instance Chapter 1 of \cite{citeulike:712984}.}, we do not expect to get insights from experiments, as it seems unlikely that any quantum gravity regime should be in reach of current experimental techniques. This means that the theory has to be built relying only on its self-consistency. That such a task should be possible at all is far from obvious, but the main surprise that came out of string theory is that consistency arguments alone allow to single out a seemingly unique (though still largely unknown) structure, M-theory. In this context, it is crucial to identify and test systematically the consistency conditions that we expect quantum theories of gravity to satisfy.

The main source of consistency conditions turns out to be the web of relations between the various partial descriptions of string theories and M-theory. Examples include the worldsheet sigma model formulations of perturbative string theories versus the target space low energy effective supergravities, the non-perturbative dualities relating the weakly coupled descriptions of M-theory and the relations between effective quantum field theories and supergravities in various dimensions obtained via compactification. Maybe the second most important class of consistency conditions arise by considerations of anomaly cancellation. In this paper, we will be interested in global gravitational anomalies, whose cancellation can be seen as a generalization to higher dimensional field theories of the modular invariance constraints familiar from 2-dimensional conformal field theories. While formulas for the global gravitational anomalies of chiral fermionic field theories have been known for a long time \cite{Witten:1985xe, MR861886}, systematic checks of global anomaly cancellation have been prevented by the lack of a general formula for the global anomaly of the self-dual field theory.

Recall that the self-dual field theory (also known as the chiral $2\ell$-form field theory) is the quantum field theory of an abelian $2\ell$-form gauge field, living on a $4\ell+2$-dimensional manifold, whose $2\ell+1$-form field strength obey a self-duality condition: $F = \ast F$. \footnote{In other dimensions, field theories cannot develop gravitational anomalies \cite{AlvarezGaume:1983ig}.} The self-dual field theory is an essential building block of the chiral effective field theories appearing in supergravity, string theory and M-theory. A few examples in various dimensions include the chiral bosons on the 2-dimensional worldsheet of the heterotic string, the chiral 2-forms in the gravity and tensor multiplets of six-dimensional supergravities, the chiral 2-forms living on several five-branes (the M5-brane in M-theory, the NS5-brane of type IIA supergravity and the NS5-brane of the $E_8 \times E_8$ heterotic string) and the Ramond-Ramond 4-form gauge field of 10-dimensional type IIB supergravity. 

In his foundational paper on global gravitational anomalies \cite{Witten:1985xe}, Witten made a proposal for the global gravitational anomaly of the self-dual field, in the case when the self-dual field has no zero modes. His motivation was to prove the absence of global gravitational anomalies in type IIB supergravity, but he succeeded to do so only when the base manifold is a ten-dimensional sphere (related to 10-dimensional Minkowski spacetime via Wick rotation). In the following, we will get a general formula and show that it implies that (a certain version of) type IIB supergravity is free of global anomalies on all ten-dimensional spin manifolds. We will also review a few other potential applications of the global anomaly formula.

%

\section{Gravitational anomalies}

We will study the anomalies of a Euclidean quantum field theory on a compact smooth spin Riemannian manifold $M$. Recall that anomalies are the breakdown of certain classical symmetries in the quantum field theory. They arise because of the lack of a path integral measure invariant under the symmetry considered, so although the classical action is invariant, the quantum partition function is not. In the case of gravitational anomalies, the classical symmetry under consideration is the group $\mathcal{D}$ of diffeomorphisms of $M$ (equivalent to the group of coordinate transformations). 

There is a geometrical way of looking at anomalies which will prove useful in the following. The metric $g$ on $M$ can be seen as an external parameter, on which the partition function $Z$ of the quantum field theory depends. Under the action of a diffeomorphism $\phi$, because of the anomaly, $Z(g)$ is not necessarily invariant. In general
\be
Z((\phi^{-1})^\ast g) = \xi(\phi, g) Z(g) \quad \xi(\phi, g) \in \mathbbm{C} \;.
\ee
It can be shown that the norm of the partition function is non-anomalous \cite{AlvarezGaume:1983ig}, so $\xi(\phi, g)$ is actually valued in $U(1)$. Consistency with the group structure requires
\be
\xi(\phi_2 \circ \phi_1, g) = \xi(\phi_2, (\phi_1^{-1})^\ast g) \xi(\phi_1, g) \;,
\ee
so $\xi$ is a 1-cocycle for $\mathcal{D}$ with value in the sheaf of $U(1)$-valued functions on $\mathcal{M}$. A 1-cocycle such as $\xi$ can be used to defines a hermitian line bundle $\mathscr{A}$ over $\mathcal{M}/\mathcal{D}$, the anomaly bundle. \footnote{$\mathcal{M}/\mathcal{D}$ is badly singular, because of the presence of metrics admitting isometries. Line bundles over this quotient are best viewed as $\mathcal{D}$-equivariant line bundle over $\mathcal{M}$. However, if we see $M$ as the one-point compactification of a non-compact Euclidean space-time, it is natural to restrict our attention to diffeomorphisms leaving the distinguished point and its tangent space fixed. With this restriction on $\mathcal{D}$,  $\mathcal{M}/\mathcal{D}$ is then smooth.} To this end, define $\mathscr{A}$ to be the quotient of $\mathcal{M} \times \mathbbm{C}$ by the equivalence relation $(g,z) \simeq (\phi(g),\xi(\phi, g)z)$. In general, the partition function is not a well-defined function on $\mathcal{M}/\mathcal{D}$, but rather a section of $\mathscr{A}$.

Except in very specific examples (like solvable two dimensional conformal field theories), checking explicitly the invariance of $Z$ under the action of the diffeomorphism group is a hopeless task. What saves the day is that it is possible to define a natural connection $\nabla_{\mathscr{A}}$ on the anomaly bundle. If the curvature and the holonomies of $\nabla_{\mathscr{A}}$ vanish, it defines a canonical trivialization of the line bundle $\mathscr{A}$, which allows one to see $Z$ as a true function over $\mathcal{M}/\mathcal{D}$ (up to an irrelevant multiplicative constant), or equivalently to show that it is invariant under the action of $\mathcal{D}$. 
In the physical terminology, the curvature of $\nabla_{\mathscr{A}}$ is known as the \emph{local anomaly} and the holonomies of $\nabla_{\mathscr{A}}$ form the \emph{global anomaly}.

It should be emphasized that an anomaly developed by a global symmetry does not indicate any inconsistency, it simply means that the symmetry is broken in the quantum field theory. Accordingly, quantum field theories developing gravitational anomalies are perfectly consistent, if not generally covariant. The troubles arise when an anomaly is developed by a symmetry which is supposed to be gauged in the full theory, as is the case of the diffeomorphism symmetry of quantum theories of gravity. In this case, we are supposed, at least formally, to perform a path integral over $\mathcal{M}/\mathcal{D}$. As it is only possible to integrate honest functions, and not sections of line bundles, the gauging of the diffeomorphism symmetry is fundamentally inconsistent if either a local or global gravitational anomaly is present. As anomalies do not depend of the scale at which the theory is probed, we obtain a consistency condition: if a quantum field theory is obtained as the low energy limit of a supposedly consistent theory of quantum gravity, its local and global gravitational anomalies have to vanish.

Gravitational anomalies typically occur in chiral fermionic theories in dimension $4\ell+2$. A chiral fermionic theory is associated to a chiral Dirac operator $D: \mathscr{S}^+ \otimes \mathscr{E} \rightarrow \mathscr{S}^- \otimes \mathscr{E}$ on $M$. Here, $\mathscr{S}^\pm$ are the even/odd (or positive/negative chirality) spinor bundle and $\mathscr{E}$ is a ``twist'' bundle with connection $\nabla_\mathscr{E}$. For instance, if the fermionic theory is coupled to a gauge field, $\mathscr{E}$ is the bundle associated to the principal bundle of the gauge field and to the representation in which the fermions transforms. If the fermionic theory is a Rarita-Schwinger theory (the gravitino of supergravities), $\mathscr{E}$ is the virtual bundle $TM \ominus 1$, where $1$ is a trivial line bundle (required to account for the presence of ghosts, see for instance \cite{AlvarezGaume:1983ig, Witten:1985xe}).

Consider $\mathcal{F} := (M \times \mathcal{M})/\mathcal{D}$, where $\mathcal{D}$ acts by its defining action on $M$ and by pull-backs on $\mathcal{D}$. $\mathcal{F}$ is a fibre bundle over $\mathcal{M}/\mathcal{D}$ with fiber $M$. Remark that each fiber is equipped with a canonical metric, defined by the projection of the fiber bundle. We get in this way a family of chiral Dirac operators over $\mathcal{M}/\mathcal{D}$. From this data, it is possible to define rigorously a line bundle $\mathscr{D}$ with connection $\nabla_{\mathscr{D}}$ over $\mathcal{M}/\mathcal{D}$, the determinant bundle of the Dirac operator \cite{MR853982, MR861886, Freed:1986hv}. Moreover, if the latter has index zero, the determinant bundle has a canonical section which can be identified with the determinant of the Dirac operator. The anomaly bundle $\mathscr{A}$ of the chiral fermionic theory coincides with $\mathscr{D}$, as a line bundle with connection.

The anomaly formulas for chiral fermionic theories describe the curvature and holonomies of $\nabla_{\mathscr{D}}$. They were obtained by Alvarez-Gaumé and Witten \cite{AlvarezGaume:1983ig} for the local anomaly and Witten \cite{Witten:1985xe} for the global anomaly. A rigorous proof and generalization was presented by Bismut and Freed in \cite{MR853982, MR861886}, using index theory techniques.

To describe local anomaly formula, remark that we can consider the bundles $TM$ and $\mathscr{E}$ as bundles over the fiber bundle $\mathcal{F}$. A less trivial fact is that they can be endowed with natural connections as bundles over $\mathcal{F}$ \cite{MR853982, MR861886}. The curvature of $\nabla_\mathscr{D}$ is then given by
\be
R_\mathscr{D} = \left[ 2\pi i \int_M \hat{A}(R_{TM}) {\rm ch}(R_\mathscr{E}) \right]^{(2)} \;,
\ee
where if $\mathscr{X}$ is a bundle with connection, $R_\mathscr{X}$ denotes its curvature, and $(\bullet)^{(2)}$ selects the 2-form component. We have used the standard notation for the A-roof genus and the Chern character:
\be
\hat{A}(R) = \sqrt{\det \frac{R/4\pi}{\sinh R/4\pi}} \;, \quad {\rm ch}(R) = {\rm Tr} \exp iR/2\pi \;.
\ee
In order to check the cancellation of local anomalies, we only have to compute the relevant polynomial in the Pontryagin classes of $TM$ and the Chern classes of $\mathscr{E}$, and check that after integration over $M$, the degree 2 part vanishes. This is equivalent to checking that the degree $4\ell+4$ part of the polynomial vanishes.

The global anomalies are more difficult to compute. We start by picking a loop $c$ in $\mathcal{M}/\mathcal{D}$. Such a loop lifts to a possibly open path in $\mathcal{M}$ whose endpoint are related by the action of a diffeomorphism $\phi_c$. We then construct the mapping torus $\hat{M}_c := \mathcal{F}|_c$, which can be alternatively defined as $\hat{M}_c = M \times I/\{(x,0) \simeq (\phi_c(x),1)\}$. Remark that each fiber $M_g$ of $\hat{M}_c$ carry a natural metric $g$. Pick a metric $g_c$ on $c$ and set $g_\epsilon = g_c/\epsilon^2 \oplus g$, a metric on $\hat{M}_c$. Consider the Dirac operator $\hat{D}_\epsilon$ on $\hat{M}_c$ twisted by $\mathscr{E}$, seen as a bundle over $\hat{M}_c$. Let $\eta_\epsilon$ be its eta invariant and $h_\epsilon$ the dimension of its space of zero modes. The holonomy of $\nabla_\mathscr{D}$ is computed by:
\be
\label{EqHolDetBM}
 {\rm hol}_{\mathscr{D}}(c) = (-1)^{{\rm index} D} \lim_{\epsilon \rightarrow 0} \exp -\pi i (\eta_\epsilon + h_\epsilon)
\ee
where we took an ``adiabatic limit'' in which the size of the base circle of $\hat{M}_c$ is sent to infinity.

Unfortunately, this formula is useless for practical checks of anomaly cancellation, because like most spectral invariants, it is hopeless to compute $\eta_\epsilon$ explicitly. Witten's remarkable idea \cite{Witten:1985xe} was to point out that if $\hat{M}_c$ bounds a spin manifold $W$, one can use the Atiyah-Patodi-Singer theorem to obtain a useful formula, of the form
\be
\label{EqGlobAnChirFerm}
\frac{1}{2\pi i} \ln {\rm hol}_{\mathscr{D}}(c) = \left({\rm index} D_W - \int_W \hat{A}(R_{TM}) {\rm ch}(R_\mathscr{E})  \right) \;.
\ee
where $D_W$ is a Dirac operator on $W$ whose restriction to the boundary $\hat{M}_c$ coincides with $\hat{D}_\epsilon$, and we left the adiabatic limit implicit. Remark that the second term is the integral of the local index density appearing in the local anomaly formula. If the local anomaly of the full theory vanishes, these terms will cancel. What remains to check is that the remaining terms add up to an integer. As ${\rm index} D_W$ is obviously an integer it might appear that no global anomaly can occur in theories free of local anomalies and built out of chiral fermionic fields. However \eqref{EqGlobAnChirFerm} is slightly sketchy and there are at least two ways in which a global anomaly can nevertheless occur. 

First, \eqref{EqGlobAnChirFerm} is valid for Weyl fermions. If the fermions are Majorana-Weyl, the right-hand side of \eqref{EqGlobAnChirFerm} should be divided by two, so a sign anomaly can occur if the index of $D_W$ is odd. Second, depending on the twist bundle $\mathscr{E}$, additional fractional factors on the right-hand side can occur. As an example, consider the case of the Atiyah-Patodi-Singer operator (the Dirac operator on $\hat{M}_c$ twisted by the spin bundle). In this case, we can take $D_W$ to be the signature Dirac operator on $W$, but it restricts to a Dirac operator on $\hat{M}_c$ whose spectrum consists of two copies of the spectrum of the Atiyah-Patodi-Singer operator. The right-hand side of \eqref{EqGlobAnChirFerm} should therefore be divided by two, as is familiar from the Atiyah-Patodi-Singer theorem for the signature operator \cite{Atiyah1973}, and a sign anomaly can occur as well.

After taking into account these details, \eqref{EqGlobAnChirFerm} solves the problem of computing gravitational anomalies for chiral fermionic theories. Unfortunately, the self-dual field theory is a chiral fermionic theory that does not fall into this framework, although its local gravitational anomaly is known to be described by half of that of chiral fermions coupled to chiral spinors (i.e. with twist bundle $\mathscr{S}^+$, the even spin bundle). In order to understand its global anomaly, we have to study its anomaly bundle.

\section{Line bundles over the space of metrics modulo diffeomorphisms}

We first have to review some facts about the topology of manifolds of dimension $4\ell+2$. The wedge product pairing is antisymmetric on $\Omega^{2\ell+1}(M)$ and endows it with a symplectic structure $\omega$. The Hodge star operator squares to $-\mathbbm{1}$ on $\Omega^{2\ell+1}(M)$, and hence defines a complex structure. Together they define a Kähler structure on $\Omega^{2\ell+1}(M)$. The Kähler structure restricts to the space of harmonic forms, and therefore to the real cohomology $H^{2\ell+1}(M,\mathbbm{R})$.

A complex structure on a $2n$-dimensional symplectic vector space such as $H^{2\ell+1}(M,\mathbbm{R})$ can be parameterized by a complex $n \times n$ matrix $\tau$ with positive definite imaginary part, belonging to a Siegel upper-half space $\mathcal{C}$. $\tau$ is defined with respect to a Darboux basis $(\{\alpha_i\},\{\beta^i\})$, such that the holomorphic tangent space is generated by the vectors $\{\alpha_i + \tau_{ij} \beta^j\}$.

In order to define the partition function of the self-dual field theory, we need an extra structure, namely a ``quadratic refinement of the intersection form'' \cite{Witten:1996hc, Witten:1999vg, Belov:2006jd}, which can be thought of as a theta function characteristic on the intermediate Jacobian $H^{2\ell+1}(M,\mathbbm{R})/H^{2\ell+1}_{\mathbbm{Z}}(M,\mathbbm{R})$, where the quotient is by the lattice of classes with integral periods. Such a characteristic can be non-canonically parameterized by $\eta \in \frac{1}{2}H^{2\ell+1}_{\mathbbm{Z}}(M,\mathbbm{R})/H^{2\ell+1}_{\mathbbm{Z}}(M,\mathbbm{R})$. This structure should be thought of as the analogue of a spin structure in the case of fermionic theories.\footnote{In dimension 2, where the self-dual field is the chiral boson, equivalent to a complex fermion, one easily check that a quadratic refinement of the intersection form is indeed a spin structure.} We will write $\mathcal{D}_\eta$ for the group of diffeomorphisms of $M$ preserving the quadratic refinement and the partition function of the self-dual field will be defined over $\mathcal{M}/\mathcal{D}_\eta$.

We already saw that Dirac operators allow to construct line bundles over $\mathcal{M}/\mathcal{D}_\eta$. In order to describe the anomaly bundle of the self-dual field theory, we however need an extra construction. A metric in $\mathcal{M}$ determines a Hodge star operator, whose restriction to $H^{2\ell+1}(M,\mathbbm{R})$ determines an element $\tau \in \mathcal{C}$. The action of $\mathcal{D}_\eta$ on $M$ induces an action on $H^{2\ell+1}(M,\mathbbm{R})$, which is symplectic with respect to $\omega$, preserves the integral cohomology and factors through a subgroup $\Gamma_\eta \subset {\rm Sp}(2n,\mathbbm{Z})$, the ``theta group''. We therefore obtain a map $\mathcal{M}/\mathcal{D}_\eta \rightarrow \mathcal{C}/\Gamma_\eta$ that we can use to pull back bundles.

The following line bundles over $\mathcal{C}/\Gamma_\eta$ will be necessary to describe the anomaly bundle. The Siegel theta constant $\theta^\eta(0,\tau)$ is a holomorphic function on $\mathcal{C}$ that can be seen as the pull back to $\mathcal{C}$ of the section of a line bundle $\mathscr{C}^\eta$ over $\mathcal{C}/\Gamma_\eta$, the theta bundle. The Hodge bundle $\mathscr{H}$ is defined by $\big(H^{2\ell+1}_{SD}(M,\mathbbm{R}) \times \mathcal{C}\big)/\Gamma_\eta$, where $H^{2\ell+1}_{SD}(M,\mathbbm{R})$ is the holomorphic subspace of $H^{2\ell+1}(M,\mathbbm{R})$. $\mathscr{K} := \det \mathscr{H}$ is its determinant bundle. Alternatively, $\mathscr{K}$ can be seen as the line bundle whose sections, when pulled-back to $\mathcal{C}$, are Siegel modular forms of weight 1 with respect to $\Gamma_\eta$.

All these bundles can be described very explicitly on $\mathcal{C}/\Gamma_\eta$ by means of factors of automorphy, which can be extracted from the functional equations that the pull-backs of their sections satisfy (see \cite{Monnier2011}).

Let us call $\mathscr{D}$ the determinant bundle of the Dirac operator coupled to chiral spinors ($\mathscr{E} = \mathscr{S}^+$, the even spinor bundle), and $\mathscr{D}_s$ the determinant bundle of the signature operator ($\mathscr{E} = \mathscr{S}$, the full spinor bundle). We endow them with their Bismut-Freed connections. One can show that $\mathscr{D}_s = \mathscr{D}^2$ as line bundles with connections. Moreover, $\mathscr{D}$ is isomorphic to $(\mathscr{K})^{-1}$ \cite{Monnier:2010ww}. We can also construct a flat bundle $\mathscr{F}^\eta := (\mathscr{C}^\eta)^2 \otimes (\mathscr{K})^{-1}$ whose holonomies are easily computed as a character $\chi^\eta$ of $\Gamma_\eta$ from the transformation formula of the Siegel theta function.

\section{The partition function and the anomaly bundle}

We now have to construct the partition function of the self-dual field and identify the bundle it is a section of. Familiar constructions such as geometric quantization \cite{Witten:1996hc, Belov:2006jd, Monnier:2010ww} or holomorphic factorization \cite{Henningson:1999dm} unfortunately cannot yield information about the global anomaly. Geometric quantization allows one to construct a (necessarily trivial) line bundle with connection over $\mathcal{M}$. This allows one to recover the local anomaly of the self-dual field theory \cite{Monnier:2010ww}, but not the global anomaly. Indeed, changes of trivializations relate (trivial) bundles on $\mathcal{M}$ that are pull-backs of bundles on $\mathcal{M}/\mathcal{D}_\eta$ differing by flat line bundles. Holomorphic factorization is useless as well: given a factorization of the partition function of the ordinary abelian gauge field on $\mathcal{M}/\mathcal{D}_\eta$, it is always possible to twist the partition function with the section of a flat bundle on $\mathcal{M}/\mathcal{D}_\eta$ and get another acceptable holomorphic factorization (see \cite{Henningson:1999dm, Monnier:2010ww} for the meaning of holomorphic in this context). 

It seems that only the path integration of a classical action allows one to extract the global anomaly. Somewhat surprisingly, constructing the partition function of the self-dual field theory from the path integration of a classical action on an arbitrary Riemannian manifold is still an open problem. Most existing actions in the literature cannot be formulated on an arbitrary Riemannian manifold \cite{Henneaux:1988gg, Pasti:1996vs}, and those for which it is possible do not seem to have the correct configuration space of fields in order for the one-loop determinant of the path integration to match what is expected from geometric quantization \cite{Belov:2006jd} and holomorphic factorization \cite{Henningson:1999dm}. See \cite{Monnier2011} for a detailed discussion. 

However, using ideas from Belov and Moore \cite{Belov:2006jd}, it is possible to construct the partition function for a pair of self-dual fields (of the same chirality) via path integration \cite{Monnier2011}. It has the form
\be
\mathcal{Z} = (\theta^\eta)^2 \cdot (\mbox{one loop determinant})
\ee
where $\theta^\eta$ is the theta constant pulled back from $\mathcal{C}/\Gamma_\eta$ to $\mathcal{M}/\mathcal{D}$ via the map from $\mathcal{M}$ to $\mathcal{C}$ reviewed above. It is possible to show that the one-loop determinant vanishes \emph{nowhere} on $\mathcal{M}/\mathcal{D}_\eta$. This is due to the fact that it is expressed in terms of determinants of Laplacians whose kernels are given by spaces of harmonic forms. The dimension of these kernels cannot jump as the metric is varied, and therefore the determinants of the Laplacians on the complement of their generic kernels never vanish. This important fact implies that the one-loop determinant is the section of a topologically trivial bundle, and we deduce that the anomaly bundle of a pair of self-dual fields $(\mathscr{A}^{\eta})^2$ is isomorphic to $(\mathscr{C}^\eta)^2$.

On the other hand, it has been known for a long time \cite{AlvarezGaume:1983ig} that the local anomaly of a pair of self-dual fields is described correctly by the curvature of the Bismut-Freed connection on $\mathscr{D}^{-1}$. We deduce that the curvatures of the connections on $(\mathscr{A}^{\eta})^2$ and on $\mathscr{D}^{-1}$ have the same local form, which in turns implies that as bundles with connections, $(\mathscr{A}^{\eta})^2$ and $\mathscr{D}^{-1}$ coincide up to a flat line bundle. These two constraints allows one to determine $(\mathscr{A}^{\eta})^2$ as a bundle with connection:
\be
(\mathscr{A}^{\eta})^2 = \mathscr{D}^{-1} \otimes (\mathscr{K})^{-1} \otimes (\mathscr{C}^\eta)^2 = \mathscr{D}^{-1} \otimes \mathscr{F}^\eta \;.
\ee
The first equality shows that $(\mathscr{A}^{\eta})^2$ is isomorphic to $(\mathscr{C}^\eta)^2$, as $\mathscr{K}$ is isomorphic to $\mathscr{D}^{-1}$. The second shows that $(\mathscr{A}^{\eta})^2$ differs from $\mathscr{D}^{-1}$ by a flat line bundle. We have therefore determined the anomaly bundle for a pair of self-dual fields and its connection \cite{Monnier2011}.

\section{The global anomaly formula}

It is now straightforward to obtain a formula for the global anomaly of a pair of self-dual fields, as the holonomies of the connection on $\mathscr{D}^{-1}$ are provided by the Bismut-Freed formula and the holonomies of $\mathscr{F}^\eta$ can be determined by the theta transformation formula. Explicitly using \eqref{EqHolDetBM} for the Dirac operator coupled to chiral spinors, 
\be
\label{EqHolSDSq}
 {\rm hol}_{(\mathscr{A}^\eta)^2}(c) = (-1)^{{\rm index} D} \lim_{\epsilon \rightarrow 0} \exp -\pi i (\eta_\epsilon + h_\epsilon) \chi^\eta(\gamma_c)
\ee
In the formula above, $\eta_\epsilon$ and $h_\epsilon$ refer to the Dirac operator $\hat{D}$ on $\hat{M}_c$ whose twist bundle is $\mathscr{S}_+$, the even spinor bundle of $M$, seen as a bundle on $\hat{M}_c$. $\chi^\eta$ is the character of $\Gamma^\eta$ describing the holonomies of $\mathscr{F}^\eta$, and $\gamma_c$ is the element of $\Gamma^\eta$ describing the action of the large diffeomorphism associated to the loop $c$ on $H_{\mathbbm{Z}}^{2\ell+1}(M, \mathbbm{R})$. As was already mentioned, formula like \eqref{EqHolSDSq} are useless for practical checks of anomaly cancellation unless they can be reexpressed in terms of a manifold $W$ bounded by $\hat{M}_c$, on the model of \eqref{EqGlobAnChirFerm}. It turns out that there is no convenient Dirac operator on $W$ which restricts to $\hat{D}$. However, by squaring \eqref{EqHolSDSq}, we can reexpress it in terms of the eta invariant of a Dirac operator $\hat{D}_s$ on $\hat{M}_c$ associated to the signature Dirac operator on $W$, originally considered by Atiyah-Patodi-Singer \cite{Atiyah1973}. We therefore obtain 
\be
\label{EqHolSD4}
 {\rm hol}_{(\mathscr{A}^\eta)^4}(c) = \exp -\pi i (\eta_0 + h_0) (\chi^\eta(\gamma_c))^2 \;.
\ee
In the equation above, we explicitly took the adiabatic limit, as the eta invariant of $\hat{D}_s$ admits a well-defined limit $\eta_0$. This squaring operation has a price, as the 2-fold ambiguity in the holonomy of $\mathscr{A}^\eta$ in \eqref{EqHolSDSq} turns into a 4-fold ambiguity. We will not be able to resolve it in a rigorous way but will perform several consistency check of our guess for the holonomy formula of $\mathscr{A}^\eta$.

The crucial step in order to rewrite \eqref{EqHolSD4} in terms of the bounded manifold $W$ is to first express it in terms of an ``Arf invariant'' of $\hat{M}_c$. We now briefly review this notion, see \cite{2012arXiv1208.1540M} for more details. The linking pairing $L$ on $H^{2\ell+2}_{\rm tors}(\hat{M}_c,\mathbbm{Z})$ admits quadratic refinements $q$. These are functions from $H^{2\ell+2}_{\rm tors}(\hat{M}_c,\mathbbm{Z})$ to $\mathbbm{Q}/\mathbbm{Z}$ satisfying:
\be
q(x+y) - q(x) - q(y) = L_G(x,y) \;, \quad q(nx) = n^2q(x) \;.
\ee
It turns out that the characteristic $\eta$ of the self-dual field determines a preferred quadratic refinement $q^\eta$ \cite{Lee1988, Monnier2011a, 2012arXiv1208.1540M}. On general grounds one can show that the argument of the Gauss sum
\be
{\rm Gauss}(q^\eta) = \sum_{g \in G} \exp 2\pi i q^\eta(g)
\ee
is a multiple of $2\pi/8$. The Arf invariant $A_\eta = A(q_\eta)$ is this argument, seen as an element of $\frac{1}{8}\mathbbm{Z}/\mathbbm{Z}$. The Arf invariant happens to satisfy the following mysterious relation \cite{Monnier2011a}:
\be
\label{EqRelArf}
(\chi^\eta(\gamma_c))^2 \exp \pi i h = \exp -\pi i 8 A_\eta(\hat{M}_c) \;.
\ee
This equation relates the Arf invariant to the number of the zero mode of $\hat{D}_s$ (expressible in terms of the cohomology of $\hat{M}_c$) and to the character $\chi^\eta$ extracted from the theta transformation formulas. We do not have a complete proof of this formula, but it is easy to check it on any given element of $\gamma_c \in \Gamma_\eta$. It would be very interesting to uncover the mathematics lurking behind it.

Combining \eqref{EqHolSD4} and \eqref{EqRelArf}, we obtain:
\be
{\rm hol}_{(\mathscr{A}^\eta)^4}(c) = \exp \pi i (\eta_0 - 8 A_\eta) 
\ee
We can now use the Atiyah-Patodi-Singer theorem \cite{Atiyah1973} to reexpress $\eta_0$:
\be
\eta_0 = \int_W L - \sigma_W \;,
\ee
where $W$ is a manifold bounded by $\hat{M}_c$, $\sigma_W$ is the signature of $W$ and $L$ is the Hirzebruch L-genus. A result of Brumfiel and Morgan \cite{Brumfiel1973} allows one to reexpress $A_\eta$ in terms of data on $W$
\be
A_\eta = \frac{1}{8}\int_W \lambda_\eta^2 - \sigma_W \quad {\rm mod 1} \;,
\ee
where $\lambda_\eta$ certain lift of the degree $2\ell+2$ Wu class of $W$ from $H^{2\ell+2}(W,\hat{M}_c,\mathbbm{Z}_2)$ to $H^{2\ell+2}(W,\hat{M}_c,\mathbbm{Z})$ that we will make more explicit in an example below. Combining these two results, we get a holonomy formula in terms of data on $W$:
\be
\label{EqHol4SD}
{\rm hol}_{(\mathscr{A}^\eta)^4}(c) = \exp \pi i \int_W (L - \lambda^2_\eta) \;.
\ee
This is a global gravitational anomaly formula for four self-dual fields of the same chirality. 

We have no way to derive from it rigorously a formula for the global anomaly of a single self-dual field, but the most naive way of taking the fourth root of \eqref{EqHol4SD} is to divide the exponent by 4: 
\be
\label{EqHol1SD}
{\rm hol}_{\mathscr{A}^\eta}(c) = \exp \frac{2\pi i}{8} \int_W (L - \lambda_\eta^2) \;.
\ee
There are several consistency checks that this formula passes. For instance, we can show that \eqref{EqHol1SD} is compatible with the local anomaly of the self-dual field theory by considering topologically trivial loops in $\mathcal{M}/\mathcal{D}$. Moreover, the holonomies of the bundles $\mathscr{A}^{\eta'} \otimes (\mathscr{A}^\eta)^{-1}$ are easy to compute: as the contribution of the one-loop determinant vanishes, they can be extracted from the theta transformation formula. Using results from Lee, Miller and Weintraub \cite{Lee1988}, one can show that \eqref{EqHol1SD} reproduces these holonomies. One can also use our knowledge of the Picard group of $\mathcal{C}/\Gamma_\eta$ \cite{2009arXiv0908.0555P} to deduce that, provided the formula above defines the holonomies of a well-defined bundle, then it is the correct one. Finally, an apparently similar formula (with some technical differences, see the introduction of \cite{2012arXiv1208.1540M}) appeared previously in the work of Hopkins and Singer \cite{hopkins-2005-70}, and can be understood as describing as well the holonomies of a line bundle.

The global gravitational anomaly formula \eqref{EqHol1SD} has a very interesting feature. In a generic theory with no local anomalies, the contribution of the other fields will cancel the first term. The second term is generically an eighth root of unity. However, as far as we are aware, chiral fermionic field theories can only contribute a sign to the global anomaly. Hence there might be situations in which the global anomaly of a self-dual field cannot be canceled by any combination of chiral fermions. 

We should mention as well that in certain six-dimensional self-dual field theories, notably the one on the worldvolume of the M5-brane, the characteristic of the self-dual field is believed to be metric-dependent \cite{Moore:2004jv}. In this case, we expect extra factors in \eqref{EqHol1SD} accounting for this metric dependence. This is currently under investigation.

\section{Type IIB supergravity}

In this section, we will check the cancellation of global gravitational anomalies in IIB supergravity. There is a subtle point concerning the topological sectors of the Ramond-Ramond gauge fields of the theory. Usually, by an abelian $p$-form gauge field, we understand a field $C$ which can be locally represented by a $p$-form, and whose field strength $H = ``dC''$ is a closed form which can have a non-trivial integral cohomology class. In more precise terms, the gauge field is a differential cohomology class (see \cite{Freed:2006yc} for a pedagogical introduction) and its topological sectors (instantons) are classified by the integral cohomology of $M$. We will call the IIB supergravity theory whose gauge fields are differential cohomology classes ``cohomological type IIB supergravity''. This is the theory that we will study here.

It should be emphasized that this theory is \emph{not} the low energy limit of the type IIB superstring. In the latter, as D-brane physics taught us \cite{Minasian:1997mm, Moore:1999gb}, the Ramond-Ramond charges and fluxes are classified by K-theory. As a result the Ramond-Ramond gauge fields are differential K-theory classes, whose instantonic configurations are classified by the K-theory of $M$. This distinction is important, because the partition function of the anomalous Ramond-Ramond four-form is given by an instanton sum. As the K-theory does not coincide with the cohomology with integral coefficients in general, the partition functions of the two theories do not coincide on certain manifolds. The extension of the global anomaly formula for self-dual fields valued in differential K-theory is an important problem currently under investigation. 

Global anomaly cancellation in Type IIB supergravity has already been studied by Witten in his original paper on global gravitational anomalies \cite{Witten:1985xe}. For the global anomaly of the self-dual field, he used the formula \eqref{EqGlobAnChirFerm} applied to the signature operator with a suitable normalization of the exponent:
\be
\label{EqHolSignOp}
{\rm hol}_{\mathscr{A}}(c) = \exp \frac{2\pi i}{8} \left(\int_W L - \sigma_W \right) \;.
\ee
Here, $W$ is again a manifold bounded by the mapping torus constructed from the loop $c$ along which we want to compute the holonomy. After combining the holonomy due to the self-dual field with the contributions of the gravitini and the dilatini, only the second term 
$\exp -\frac{2\pi i}{8} \sigma_W$ remains.

Let us stress that this is a frightening result, as a priori there is no reason for the signature of $W$ to be a multiple of 8. In \cite{Witten:1985xe}, Witten managed to show that when $M$ is a 10 sphere, $\sigma_W$ is indeed a multiple of 8. He also warned that his result should be trusted only when $H^{5}(M,\mathbbm{Z}) = 0$. In retrospect, we see that in this case, the zero modes of the self-dual field do not contribute and we have $(\mathscr{A})^4 = (\mathscr{D}_s)^{-1}$, justifying the use of the holonomy formula for the determinant bundle of the signature operator (with the exponent suitably divided by 4).

Let us compare this result with what can be obtained from our formula. The anomaly formula \eqref{EqHol1SD} is useful only if we can determine the cohomology class $\lambda_\eta$ for the physically relevant choice of characteristic $\eta$. Fortunately, in the case of 10-dimensional type IIB supergravity, $\lambda_\eta$ vanishes \cite{Witten:1996hc, Witten:1999vg}. 

For $\lambda = 0$, \eqref{EqHol1SD} predicts
\be
{\rm hol}_{\mathscr{A}}(c) = \exp \frac{2\pi i}{8} \int_W L 
\ee
Comparing with \eqref{EqHolSignOp}, we see that the term involving $\sigma_W$, which was apparently signaling an anomaly, is absent and the global anomaly vanishes identically. Therefore Type IIB supergravity, at least in its ``cohomological'' flavor, is free of global gravitational anomalies.

One may wonder how this can be compatible with Witten's result. In fact, results of Brumfiel and Morgan \cite{Brumfiel1973} allow to show that the signature of $W$ is always a multiple of $8$ when $M$ has no middle-degree cohomology \cite{Monnier2011a}. Witten's formula is therefore correct when $H^{5}(M,\mathbbm{Z}) = 0$, it simply does not make the anomaly cancellation manifest.\\

There are several other check of global gravitational anomaly cancellation which need to be performed. We already mentioned the extension of the check above to ``K-theoretical'' type IIB supergravity. It would be important to derive a global anomaly formula for the various five-branes, as it could provide some non-trivial constraints on five-brane instantons, which are of great importance in phenomenological string models. Global gravitational anomalies could also provide new constraints on six-dimensional supergravities, with applications to the six dimensional landscape. We hope that the systematic analysis of the constraint of global anomaly cancellation will provide new insights in these subjects.

\providecommand{\href}[2]{#2}\begingroup\raggedright\endgroup

%

\end{document}